\pdfoutput=1
\documentclass{article}
\usepackage[a4paper, total={6in, 8in}]{geometry}
\usepackage[utf8]{inputenc}
\usepackage[T1]{fontenc}
\usepackage{authblk}
\usepackage{amsmath}
\usepackage{amssymb}
\usepackage{physics}
\usepackage{graphicx}
\graphicspath{{img/}}
\usepackage{hyperref}
\usepackage{subcaption}

\hypersetup{hidelinks}

\DeclareMathOperator{\epstrot}{\varepsilon_{\mathrm{trott}}}

\title{phase2: Full-State Vector Simulation \\ of Quantum Time Evolution at Scale}

\author[1]{Marek Miller\thanks{Corresponding author: \href{mailto:mlm@math.ku.dk}{mlm@math.ku.dk}}}
\author[1]{Jakob G\"unther}
\author[1,2]{Freek Witteveen}
\author[3,4]{Matthew S. Teynor}
\author[5]{Mihael Erakovic}
\author[5]{Markus Reiher}
\author[3,4]{Gemma C. Solomon}
\author[1]{Matthias Christandl}

\affil[1]{\footnotesize Quantum for Life Centre, Department of Mathematical Sciences, University of Copenhagen, Copenhagen, Denmark}
\affil[2]{\footnotesize QuSoft and CWI, Amsterdam, The Netherlands}
\affil[3]{\footnotesize Nano-Science Center and Department of Chemistry, University of Copenhagen, Copenhagen, Denmark}
\affil[4]{\footnotesize NNF Quantum Computing Programme, Niels Bohr Institute, University of Copenhagen, Copenhagen, Denmark}
\affil[5]{\footnotesize Institute for Molecular Physical Science, Department of Chemistry and Applied Biosciences, ETH Z\"urich, Z\"urich, Switzerland}

\date{\small July 14, 2026\\[2ex]
\normalsize Author's version of: Commun.\ AI Comput.\ \textbf{1}, 5 (2026),
\href{https://doi.org/10.1038/s44488-026-00002-2}%
{doi:10.1038/s44488-026-00002-2}}

\begin{document}

\maketitle

\begin{abstract}
Classical simulation of quantum computers is essential for designing and benchmarking quantum algorithms.
Here we present \texttt{phase2},
a full-state-vector simulator optimised for sequences of many-qubit Pauli rotations on distributed CPU and GPU clusters.
Exploiting the common-suffix structure of Pauli-rotation circuits,
the implementation reduces inter-node communication and achieves two orders of magnitude speedup for grouped rotations.
We demonstrate weak and strong scaling to 40 qubits across $512$ NVIDIA H100 GPUs using $32$\,TB of distributed memory.
Applying the simulator to Hamiltonian time evolution of ruthenium-ligand active spaces up to 40 qubits,
we find that the empirical Trotter error lies more than two orders of magnitude below the rigorous analytic upper bound for every active space in which the fit converged (up to 32 qubits).
Practical circuit depths and simulation costs are therefore substantially smaller than the conservative estimates suggest.

\end{abstract}

\label{sec:intro}

Classical simulation of quantum computers plays two practical roles in the development of quantum algorithms.
It produces numerically exact reference data against which approximate quantum or classical methods can be validated,
and it allows algorithm designers to measure empirical performance at problem sizes where analytic guarantees do not apply.
The algorithms most relevant to near-term hardware,
including variational methods and Hamiltonian time evolution,
generally lack tight accuracy bounds;
their behaviour must be characterised numerically,
exactly the regime in which the simulation itself becomes computationally demanding.

Several paradigms address this demand by trading exactness for efficiency.
Tensor-network methods \cite{vidal2003efficient,markov2008simulating,pan2022simulation,ayral2023density,tindall2024efficient} exploit limited entanglement.
Stabiliser decompositions \cite{bravyi2019simulation} are efficient for Clifford-dominated circuits.
Pauli propagation \cite{beguvsic2024fast,beguvsic2025real} truncates sparse operator representations.
Each approach restricts the states or gates it can represent exactly;
see \cite{xu2023herculean} for a comparative review.
Full-state vector simulation makes no such restriction:
all $2^n$ amplitudes are held in memory and updated at every step.
Memory scales exponentially with qubit count,
but the representation is exact to machine precision ---
the property that makes state-vector simulation the reference against which other methods are calibrated.
Established implementations include Qiskit \cite{qiskit2024} and NVIDIA cuQuantum \cite{bayraktar2023cuQuantum};
see \cite{jamadagni2024benchmarking} for a recent benchmark.

This paper reports \texttt{phase2},
a full-state-vector simulator specialised for Hamiltonian time evolution:
given a Hamiltonian $H$ and state $|\psi\rangle$,
it computes $\langle \psi | e^{itH} | \psi \rangle$ through a distributed implementation of three algorithms \cite{gunther2025phase} ---
the deterministic Trotter product formula,
qDRIFT,
and a partially randomised scheme (see \nameref{sec:algorithm_desc}).
The gate set is restricted to \emph{Pauli rotations} $\exp(i\phi P)$.
One- and two-qubit Pauli rotations form a universal gate set,
and high-weight rotations arise directly in electronic-structure Hamiltonians \cite{peruzzo2014variational,wiersema2020exploring,grimsley2019adaptive} and in the quantum approximate optimisation algorithm \cite{blekos2024review},
so the specialisation covers the circuit class most relevant to chemistry and combinatorial optimisation.

Characterising the Trotter error motivates a second line of work in the paper.
Product-formula implementations of Hamiltonian simulation introduce a discretisation error whose rigorous upper bounds tend to be loose for realistic systems \cite{poulinTrotterStepSize2015, childsFirstQuantumSimulation2018, martinez-martinezEstimatingTrotterApproximation2024, reiherElucidatingReactionMechanisms2017};
the gap between bound and observation can span several orders of magnitude.
Direct measurement at chemistry-relevant scale has been limited by simulator capacity,
and because product-formula circuit depth grows with the number of Hamiltonian terms,
large chemistry Hamiltonians are tackled here with the partial randomisation scheme of \cite{gunther2025phase}.
Prior state-vector simulations at 40 qubits exist \cite{willsch2023large,guerreschi2020intel};
the present work combines Pauli-rotation-optimised kernels with distributed execution to characterise Trotter-error scaling for a series of ruthenium-ligand active spaces up to 40 qubits under a non-standard Hamiltonian partitioning.

The remainder of this paper is organised as follows.
The Results section describes the simulator implementation,
reports its weak- and strong-scaling behaviour on CPU and GPU clusters,
benchmarks it against an established reference,
and presents the Trotter-error analysis.
The discussion interprets the findings,
addresses limitations and outlines future directions.
The Methods section provides the mathematical model,
the quantum algorithms,
the Hamiltonian preparation,
the data-analysis protocol and the benchmarking methodology.

\section*{\label{sec:Results}Results}

The method described in \nameref{sec:software_impl} simulates exactly the application of sequences of Pauli rotations on large quantum systems.
We demonstrate effective scaling on CPU and GPU clusters (see \nameref{sec:scalability}),
as well as its performance on grouped common-suffix Pauli rotations,
which in our benchmarks exceeds an established reference by up to two orders of magnitude for large common-suffix group sizes ($L=100$);
for individual rotations ($L=1$), the speedup is $1$--$10\times$ (see \nameref{sec:performance}).
We conducted a numerical experiment to examine the Trotter error of the time evolution given by the Hamiltonian representing the ruthenium complex NKP-1339 (see \nameref{sec:experimental_setup} and \nameref{sec:hamiltonian_prep}).
The section \nameref{sec:simulation} presents in detail the results of the computation.

\subsection*{\label{sec:software_impl}Software implementation}
Our method employs a straightforward simulation technique in which the quantum state $|\psi\rangle$ of an $n$-qubit system is represented as a contiguous array of $2^{n}$ complex probability amplitudes.
The simulation is exact,
but for more than $n=30$ qubits, the memory and compute requirements exceed a single server.
We give an optimised implementation for applying circuits consisting of Pauli rotations $\exp(i \phi \tilde{P})$,
and use the simulator for deterministic (Trotter) and (partially) randomised product formulas for Hamiltonian simulation \cite{gunther2025phase}.
Here, the time evolution along a Hamiltonian $H = \sum_l \alpha_l \tilde{P}_l$ is approximated by a sequence of Pauli rotations $\tilde{P}_l$ over small angles.
See~\nameref{sec:algorithm_desc} below for an outline of the quantum algorithms simulated in this study.

Using various optimisation strategies, which we discuss in detail below,
we simulated Hamiltonian time evolution of a chemically relevant 40-qubit system on a multi-GPU cluster.
The 40-qubit run uses $32$\,TB of distributed memory across $512$ NVIDIA H100 GPUs;
a companion 36-qubit run uses $1$\,TB across 16,384 CPU cores.
The technique is a special case of the Schr\"odinger algorithm for simulating quantum computers \cite{raedt2019massively,wu2019full};
a formal model is given in \nameref{sec:simul_model}.

Starting from the initial state $|\psi\rangle$,
the simulation algorithm proceeds by applying the sequence of the unitary operators
$e^{i \varphi \tilde{P}}$ on $|\psi \rangle$,
where $\varphi \in \mathbb{R}$,
and
$\tilde{P} = P_{1} \otimes P_{2} \otimes \ldots \otimes P_{n}$,
$P_{k} \in \left\{ I, \sigma_{x}, \sigma_{y}, \sigma_{z} \right \}$,
$k = 1, 2, \ldots n$.
We call the operator $\tilde{P}$ a \emph{Pauli string},
and the unitary $e^{i \varphi \tilde{P}}$,
a \emph{Pauli rotation}.
Note the order of qubits from the least to the most significant one.
Observing that for every Pauli string $\tilde{P}$,
$e^{i \varphi \tilde{P}} = \cos \varphi \, I + i \sin \varphi \, \tilde{P}$,
we can effect the action of the unitary on the array of amplitudes in two steps by
(i) a transformation,
given by $\tilde{P}$,
similar to transposition of the array,
and (ii) by taking a linear combination of elements.
This is a consequence of the particular structure of $\tilde{P}$ as a direct sum of 2-by-2 matrices,
for all $n$ and the Pauli operators $P_{1}, P_{2}, \ldots, P_{n}$.
(See \nameref{sec:simul_model} for further details,
as well as for proofs of other mathematical facts stated in this section.)

Drawing from previous work on high-performance simulation of quantum circuits \cite{jones2023distributed,da2020qsystem,loizeau2025quantum},
notably the QuEST software library \cite{jones2019quest},
we developed a system geared towards a large-scale distributed workload,
with the representation of the quantum state array divided evenly between a power-of-two number of concurrent processes (\emph{MPI tasks}).
We copy QuEST's approach to efficient synchronisation between the tasks by keeping buffers of size equal to the size of the local partition of the distributed state.
Although this method effectively doubles the amount of memory required for the simulation,
it does not increase overall data transfer and simplifies the addressing of the elements of the distributed array.
This, in turn, makes it possible for multiple unitaries of the form $e^{i \varphi \tilde{P}}$ to be applied in a single simulation step,
see Eq.\,\eqref{eq:core} below.

We then optimise the algorithm by observing that:
(i)~every Pauli string $\tilde{P} =  P_{1} \otimes P_{2} \otimes \ldots \otimes P_{n}$,
for $n \leq 64$,
can be efficiently represented as two 64-bit integers in such a way that the result of the action of $\tilde{P}$ on a computational basis state $|i\rangle$,
$i=0,1,\ldots 2^{n} -1$,
is obtained by using only a minimal number of logical instructions: AND,
XOR and bit parity.
This effectively makes processing elements of the array of amplitudes as fast as memory access time.
(ii)~The particular form of the unitary $e^{i \varphi \tilde{P}}$ as a direct sum of 2-by-2 matrices reduces the data transfer required for distributed simulation to pairwise exchanges between MPI tasks,
occurring simultaneously for all tasks.
For a multi-GPU simulation,
the same matrix structure enables the execution of the entire unitary in parallel using compact CUDA kernels.
This eliminates the need for global synchronisation or locking to maintain coherence of the distributed state.
(iii)~In order to minimise further the data transfer,
we group similar Pauli rotations using the following identity:
\begin{equation}
\label{eq:core}
\prod_{l=1}^{L} e^{i \varphi_l \tilde{P}_l \otimes \tilde{Q}} =  \prod_{l=1}^{L} e^{i \varphi_l \tilde{P}_l} \otimes
\frac{I + \tilde{Q}}{2} +
\prod_{l=1}^{L} e^{-i \varphi_l \tilde{P}_l} \otimes
\frac{I - \tilde{Q}}{2},
\end{equation}
where $\varphi_{l} \in \mathbb{R}$,
$\tilde{P}_{l}$ and $\tilde{Q}$ are Pauli strings,
$l = 1, 2, \ldots, L$.
If two terms,
say
$e^{i \varphi_{1} \tilde{P}_{1} \otimes \tilde{Q}}$ and
$e^{i \varphi_{2} \tilde{P}_{2} \otimes \tilde{Q}}$,
share a common suffix $\tilde{Q}$ of length $m$,
such that $2^{m}$ is the number of MPI tasks participating in the computation,
only one pairwise exchange of local array partitions is needed to achieve the action of the concatenated unitary
$e^{i \varphi_{1} \tilde{P}_{1} \otimes \tilde{Q}} e^{i \varphi_{2} \tilde{P}_{2} \otimes \tilde{Q}}$ on the distributed state.
The empirical distribution of group sizes for the ruthenium-ligand Hamiltonian studied here is discussed below.
This approach reduces the data transfer overhead and hence the overall simulation time.
See also \nameref{sec:performance} section below.

\texttt{phase2} runs on CPU and GPU clusters in a standard HPC environment.
At its core is a C library that implements the three algorithms of \cite{gunther2025phase} on a distributed state array.
The command-line driver \texttt{ph2run} coordinates the computation and handles input and output;
a Python wrapper exposes the library for scripted use.
Both the library and the associated application are implemented in standard C,
utilising \texttt{MPI} \cite{mpi41} for inter-process communication and \texttt{HDF5} \cite{koziol2014hdf5} for managing parallel I/O.
The GPU version of the library employs custom CUDA kernels,
which were compiled,
optimised,
and linked with \texttt{nvcc} for the NVIDIA Hopper architecture: \texttt{sm\_90a}.
For GPU computation,
our strategy involves a single MPI task per GPU,
with one CPU core acting as a host that controls the CUDA device,
and GPU devices communicating via MPI directly using the CUDA Unified Memory Model \cite{cuda2024guide}.
We automate the computation through standard Unix shell scripting and \texttt{Makefiles}.

\begin{figure}
\centering
\includegraphics[width=0.5\linewidth]{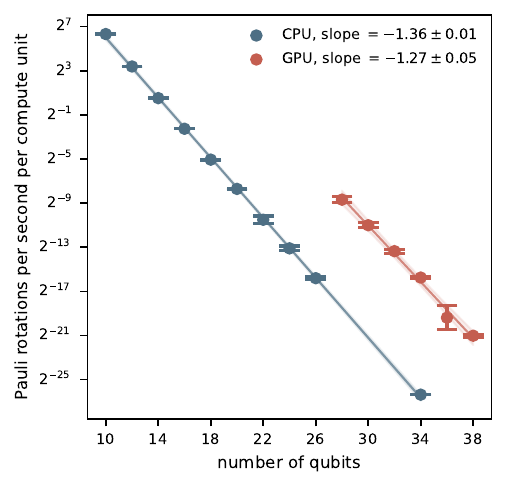}
\caption{Weak scaling of the simulation.
Per-compute-unit throughput (Pauli rotations per second) versus number of qubits $n$ on a $\log_{2}$ scale.
Blue: CPU (Euler,
AMD EPYC 9654,
compute unit = one core);
red: GPU (Gefion,
NVIDIA H100,
compute unit = one device).
Points are geometric means over SLURM jobs;
error bars show $\pm 1$ s.d.\ of $\log_{2}$ throughput.
Lines: log-linear fits with $95\%$ confidence bands.
Fitted slopes: $-1.36 \pm 0.01$ (CPU),
$-1.27 \pm 0.05$ (GPU); ideal slope is $-1$.}
\label{fig:weak-scaling}
\end{figure}

\subsection*{\label{sec:scalability}Scalability}
The algorithm exhibits effective weak and strong scaling \cite{amdahl2007validity,gustafson1988reevaluating}.
Fig.~\ref{fig:weak-scaling} demonstrates weak scaling of the implementation.
The simulation speed,
measured as the average number of Pauli rotations per second per \emph{compute unit}
(one CPU core for the CPU cluster and one GPU device for the GPU cluster),
decreases proportionally to the size of the distributed quantum state,
which scales like $2^{n}$,
where $n$ is the number of qubits.
For each SLURM job,
the per-compute-unit throughput is taken as the total number of Pauli rotations executed divided by the total wall-time and by the number of compute units allocated to the job.
The state-distribution technique scales up to 38 qubits with the slopes shown in Fig.~\ref{fig:weak-scaling}.
Simulation speed varies between runs due to (i)~Hamiltonian structure,
(ii)~the choice of simulation parameters,
and (iii)~cluster interconnect speed and node occupancy.

Strong scaling is reported in Fig.~S1 of the Supplementary Material:
for a fixed size of the computational task,
determined by the size of the distributed quantum state \emph{and} the total number of Pauli rotations,
increasing the number of CPU cores or GPU devices results in a nearly linear speed-up on the log-log scale,
indicating that computational resources are allocated and utilised efficiently.

\subsection*{\label{sec:performance}Performance}
Full-state-vector simulation of Hamiltonian time evolution is \emph{memory-bound}:
for $n \geq 20$, the state vector already exceeds CPU cache,
and memory access time dominates compute
(see the single-CPU comparison with QuEST\,v3 and Qiskit Aer in Fig.~S2 of the Supplementary Material).
Beyond $n = 36$ the state exceeds $1$\,TB and no longer fits on a single server;
the problem becomes \emph{distributed-memory bound},
and network bandwidth joins memory bandwidth as a primary cost.
On a single NVIDIA H100 GPU at $n = 28$ qubits, the partially randomised algorithm sustains $169$ Pauli rotations per second,
corresponding to a memory throughput of $1.32$\,TB/s on the $8$\,GB state ---
close to the device's $3.35$\,TB/s bandwidth limit quoted by the manufacturer.

\begin{figure}
\centering
\includegraphics[width=\linewidth]{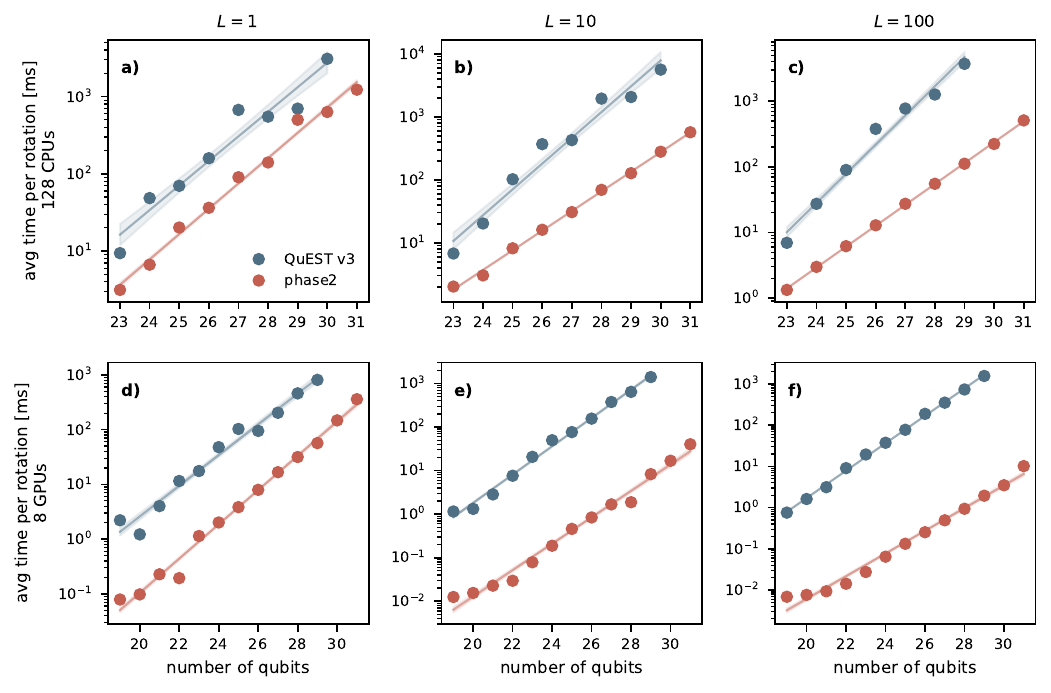}
\caption{Performance comparison: \texttt{phase2} (red) versus QuEST\,v3 (blue).
Average time per Pauli rotation (ms) versus number of qubits for group sizes
$L = 1$ (\textbf{a},\,\textbf{d}),
$10$ (\textbf{b},\,\textbf{e}),
$100$ (\textbf{c},\,\textbf{f}).
Top row: 128 CPU cores (Intel Xeon Platinum 8480C); bottom row: 8 NVIDIA H100 GPUs.
Lines: log-linear fits; bands: $95\%$ confidence intervals.}
\label{fig:ph2-quest-bench}
\end{figure}

We compare performance against QuEST\,v3,
chosen because (i)~our design was directly inspired by QuEST,
(ii)~QuEST is among the fastest distributed simulators \cite{jamadagni2024benchmarking},
and (iii)~it supports the same HPC execution environment.
Our benchmark involves measuring the time needed to perform a group of $L=1$,
$10$,
or $100$ Pauli rotations that share a common suffix $\tilde{Q}$ as in Eq.~\eqref{eq:core},
for the number of qubits $n=16, 17, \ldots, 31$,
and the number of CPU cores: $1$,
$8$ and $128$.
We also performed an analogous measurement for the GPU-based computation,
with the number of devices: $1$ and $8$.
Refer to \nameref{sec:benches} for a detailed comparison summary.
Fig.~\ref{fig:ph2-quest-bench} reports the speedup of \texttt{phase2} over QuEST\,v3 across all tested configurations.
Speedup grows with workload: individual rotations ($L=1$) run $1$--$10\times$ faster,
and grouped rotations with $L=100$ exceed $100\times$.

\subsection*{\label{sec:experimental_setup}Experimental setup}
The numerical calculation to estimate the Trotter error constant for a Hamiltonian model of a ruthenium ligand NKP-1339 in a drug-protein complex \cite{Trondl2014, Peti1999},
was performed between January and March 2025 in the two HPC facilities:
(i) a CPU cluster Euler at ETH Z\"urich,
and
(ii) a NVIDIA GPU cluster Gefion,
operated by Danish Centre for AI Innovation (DCAI) and located in Copenhagen.
On Euler,
the simulation of a system of 36 qubits used up to 16~384 CPU cores (AMD EPYC 9654 CPU) and 8\,TB of memory,
distributed among 64 nodes.
The Gefion architecture consists of hybrid servers (NVIDIA DGX nodes) equipped with 2 Intel Xeon Platinum 8480C processors and 8 NVIDIA H100 GPUs.
For our simulation,
we used up to 64 DGX nodes (512 GPUs)
and 32\,TB of GPU memory to simulate the time evolution of a 40-qubit Hamiltonian.

Euler was used for calibration and runs up to 36 qubits;
Gefion for the main 40-qubit experiment.
The simulation work was divided into several hundred jobs that typically lasted up to 24~h.
The complete calculation,
including the testing and deployment phase,
took approximately $1.33 \times 10^{7}$\,CPU\,h and $5 \times 10^{4}$\,GPU\,h.

\subsection*{\label{sec:simulation}Simulation}

The simulator is applied to estimate the Trotter-error scaling for product-formula Hamiltonian simulation,
where $N$ spatial orbitals map to $2N$ qubits.

\begin{figure}
    \centering
    \includegraphics[width=\linewidth]{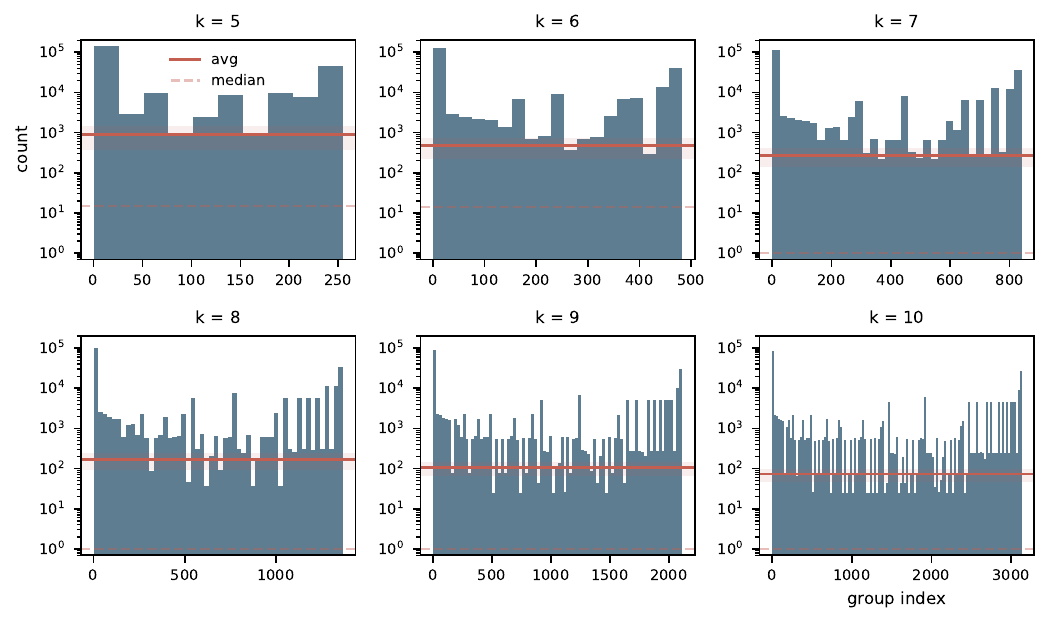}
    \caption{Common-suffix group-size distribution for the ruthenium-ligand Hamiltonian at $n = 40$ qubits,
    one panel per suffix length $k = 5$--$10$.
    x-axis: lexicographically sorted group index;
    y-axis (log scale): number of Hamiltonian terms per group.
    Solid line: mean; dashed line: median; band: $\pm 1$ s.e.}
    \label{fig:ruth-hist}
\end{figure}

For the numerical results, we use a Hamiltonian modelling a ruthenium ligand complex (see section \nameref{sec:hamiltonian_prep}).
When performing a Trotter-product formula (Eq.~\eqref{eq:define trotter hamiltonian}) or the deterministic part of the partially randomised algorithm \cite{gunther2025phase},
the Hamiltonian terms are reordered lexicographically so that Pauli rotations sharing a common suffix are grouped via Eq.~\eqref{eq:core}.
Fig.~\ref{fig:ruth-hist} shows the resulting group-size distribution at $n=40$ qubits.
At $k=9$ (the partitioning used for 512 GPUs),
groups of size $L \geq 100$ are common;
the average group size is $108$ (Table~S1).
Fig.~S3 of the Supplementary Material shows how average and median group size scale with the number of qubits.

\begin{figure}
    \centering
    \includegraphics[width=0.5\linewidth]{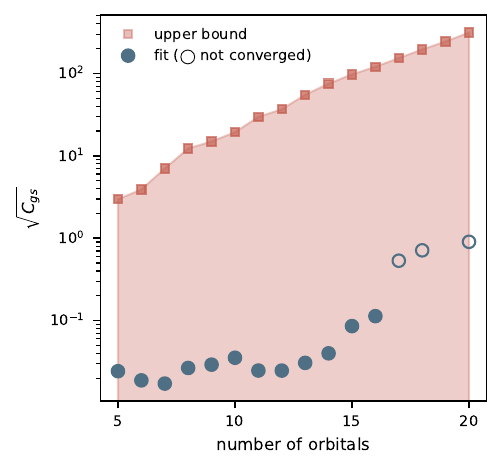}
    \caption{Trotter-error prefactor $\sqrt{C_{gs}}$ versus number of spatial orbitals
    (qubits $= 2 \times$ orbitals),
    log scale.
    Filled blue circles: converged numerical fits;
    hollow blue circles: fits that did not fully converge.
    Red squares with band: analytical upper bound from Hamiltonian coefficients.}
    \label{fig:trotter-error-scaling}
\end{figure}

To deal with the large number of Hamiltonian terms in the most efficient way,
we use a partial randomisation scheme \cite{gunther2025phase,gunther2025use}.
We simulate quantum phase estimation to reproduce the ground state energy as it would be computed by a quantum computer.
Fig.~\ref{fig:trotter-error-scaling} presents one of the main results of the simulation:
the scaling of the Trotter error with system size,
up to 32 qubits (16 orbitals).
This scaling is captured by the Trotter-error prefactor $C_{gs}$,
defined by $\epstrot = C_{gs}\,\delta^{2}$ with $\delta$ the Trotter step size (see \nameref{sec:algorithm_desc}).
For quantum algorithms using product formula simulations,
the cost scales with $\sqrt{C_{gs}}$.
The numerically computed values of $\sqrt{C_{gs}}$ are one to two orders of magnitude below the analytical upper bound (Fig.~\ref{fig:trotter-error-scaling}).
$C_{gs}$ increases with system size but less regularly than the upper bound.

We also performed simulations for larger system sizes,
corresponding to 17,
18,
19 and 20 orbitals,
or 34,
36,
38 and 40 qubits,
respectively.
For 34--40 qubits,
precision was insufficient to extract reliable $C_{gs}$ values,
but time evolution using the partially randomised product formula could still be performed.
The 40-qubit simulations performed $1.26 \times 10^{4}$ Pauli rotations on the full 512-GPU configuration.
Fig.~S4 of the Supplementary Material shows the resulting amplitude for short time values;
the signal behaviour is consistent with a correct implementation,
providing a qualitative check at the largest system size.
See \nameref{sec:data_analysis} for details.

\section*{\label{sec:Discussion}Discussion}
We have developed and optimised a software package for exact full-state-vector simulation of Hamiltonian time evolution,
targeting distributed CPU and GPU clusters.
Scalability has been demonstrated up to 38 qubits,
the largest system size the available resources support.
Performance is strongest in the grouped-rotation regime with large common-suffix group size $L$ (see \nameref{sec:performance}).

In \cite{jones2023distributed} the benefit of expressing the Pauli rotation as a direct sum of linear transformations for distributed computing was discussed.
This property is one of the key optimisations.
However,
the main speed-up of our implementation was obtained by combining it with the approach based on Eq.~\eqref{eq:core},
together with an efficient multi-GPU implementation.
Software implementations that lack the common-suffix optimisation of Eq.~\eqref{eq:core} require a separate MPI exchange and CUDA synchronisation step per Pauli rotation,
which accounts for the observed performance gap.
An analysis of performance gains from grouping Pauli strings in the context of expectation-value evaluation was reported in \cite{teranishi2024lazy};
a related technique has been implemented in QuEST\,v4 (January 2025).

The observed Trotter error is one to two orders of magnitude below the theoretical upper bounds.
This discrepancy suggests that practical cost may be lower than conservative estimates predict.
Additionally,
when considering future simulations,
it may be beneficial to adopt an inverse approach: by conducting simulations with a greater permissible error,
one could align with the theoretical limits while incurring lower simulation cost.

Several limitations should be noted.
Full-state-vector simulation scales exponentially in memory ---
our 40-qubit runs consumed 32\,TB,
and each additional qubit doubles the requirement.
The implementation handles only Pauli-rotation circuits;
arbitrary gate sets must first be compiled into Pauli rotations,
potentially increasing circuit depth.
Hybrid CPU/GPU execution within a single DGX node did not improve performance:
the computation is network-bandwidth-bound,
so additional compute units on the same node yield no speedup.

The common-suffix optimisation is Hamiltonian-dependent:
random Hamiltonians would yield small groups and little speedup,
whereas molecular Hamiltonians exhibit the heavy-tailed group-size distribution shown in Fig.~\ref{fig:ruth-hist} that the optimisation exploits.
For the ruthenium-ligand Hamiltonian at $n=40$,
$k=9$,
the largest group ($\tilde{Q} = I^{\otimes 9}$,
86,886 terms) is trivially parallelisable ---
it acts as the identity on the upper qubits and requires no inter-node communication.
The next 18 groups,
each containing ${\sim}4900$ terms with non-trivial suffixes, such as $ZZZZZZXII$ and $YIIIIIIII$,
collectively account for a further 38.5\% of the 229,140 Hamiltonian terms (Table~S1).
These groups require the same single pairwise MPI exchange as any other grouped rotation (Eq.~\eqref{eq:core}),
yet amortise that cost over thousands of terms.
The top 19 groups cover 76.4\% of all terms;
the average group size at $k=9$ is 108.
Without grouping,
each of the 229,140 terms would require a separate communication step.
The speedup is therefore a consequence of the heavy-tailed distribution,
not an artefact of the identity group alone.

The Trotter-error scaling (Fig.~\ref{fig:trotter-error-scaling}) was characterised for a single molecular system up to 32 qubits;
for larger active spaces (34--40 qubits) the numerical precision was insufficient to extract reliable $C_{gs}$ values.
Generalisation to other molecular systems remains open.
Since the computation is memory-bound,
a natural next step is to pipeline memory access and overlap communication with computation,
executing multiple Pauli rotations per pass over the amplitude array \cite{ramamoorthy1977pipeline}.
Efficient pipelining of non-local access to a distributed array remains an open challenge.

Full-state-vector simulation is needed to obtain numerically exact reference data for validating quantum algorithms applied to chemistry.
The specialised implementation reduces simulation time by a factor of ${\sim}100$ for the 40-qubit ruthenium-ligand Hamiltonian,
making this regime feasible on current HPC hardware.

\section*{\label{sec:methods}Methods}

\subsection*{\label{sec:simul_model}Mathematical model}

The state $|\psi\rangle$ of an $n$-qubit system is stored as an array of $2^{n}$ complex amplitudes (\texttt{\_Complex double}),
requiring $2^{n+4}$ bytes.
The array is divided equally among $M = 2^{m}$ MPI tasks,
one per CPU core or GPU device,
each holding a partition $A_k$ of $N = 2^{n-m}$ amplitudes and an auxiliary buffer $B_k$ of the same size,
for a total memory footprint of $2^{n+5}$ bytes.

This partitioning separates the Hilbert space into $n{-}m$ ``lower'' qubits (local to each task) and $m$ ``upper'' qubits (indexed by the task rank $k$).
A unitary $U \otimes I_m$ acting only on lower qubits is applied independently by each task to its partition --- no communication is required.
A Pauli string $\tilde{P}$ acting on the upper qubits induces a bijection $k \mapsto k'$;
implementing $I_{n-m} \otimes \tilde{P}$ therefore reduces to a single pairwise exchange of partitions $A_k$ and $A_{k'}$.
See Supplementary Methods for the full derivation and \cite{jones2023distributed} for an alternative treatment.

When a sequence of $L$ Pauli rotations shares a common suffix $\tilde{Q}$ on the upper qubits (Eq.~\eqref{eq:core}),
the entire group requires only one buffer exchange:
the implementation loads $|\psi'\rangle = (I_{n-m} \otimes \tilde{Q})|\psi\rangle$ into the buffers,
forms the superpositions $(|\psi\rangle \pm |\psi'\rangle)/\sqrt{2}$,
and applies the $L$ rotations $\prod_l e^{\pm i\varphi_l \tilde{P}_l}$ locally on lower qubits.
The proof follows by induction from the spectral decomposition of Pauli strings (Supplementary Methods).

Pauli strings are encoded as two 64-bit integers,
so that the action $\tilde{P}|i\rangle = \omega|j\rangle$ reduces to a bitwise XOR ($j = p_1 \oplus i$) and a parity computation for the phase $\omega \in \{1, i, -1, -i\}$,
achieving memory-access-limited throughput (see Supplementary Methods for details).

\subsection*{\label{sec:algorithm_desc}Hamiltonian simulation}
The algorithm tested here computes molecular ground-state energies on a quantum computer by combining Robust phase estimation (RPE) with Hamiltonian simulation via partially randomised product formulas \cite{gunther2025phase}.
The integrated analysis in \nameref{sec:data_analysis} tightens error bounds and the derived resource estimates relative to prior work,
and circuit depth decreases for the high-overlap initial states typical of chemistry problems \cite{gunther2025phase}.
Classical simulation of molecular electronic-structure Hamiltonians under partially randomised product formulas has not been reported before.

Consider a Hamiltonian of the form
\begin{align}
    H = \sum_{l=1}^L h_l \tilde{P}_l,
\end{align}
with $h_l\in\mathbb{R}$,
the $\tilde{P}_l$ tensor products of Pauli operators on $N$ qubits,
and $\lambda = \sum_{l=1}^L \abs{h_l}$.
Let $\ket{\psi_l}$ denote the $l$-th eigenstate of $H$,
and take $\ket{\psi}$ to be an initial state with ground-state overlap $\abs{\bra{\psi_0}\ket{\psi}}^2 \geq \eta$ for some $\eta \in (0,1]$.
The task is to find the ground-state energy $E_0$ to precision $\varepsilon>0$.
Phase estimation solves it at the optimal $\mathcal{O}(\varepsilon^{-1})$ (Heisenberg) scaling in total evolution time.

Time evolution is discretised into Trotter steps of size $\delta$,
\begin{align}\label{eq:define trotter hamiltonian}
    e^{i\delta H} \approx \prod_{l=1}^L e^{i\delta h_l\tilde{P}_l} = e^{i\delta \tilde{H}(\delta)}.
\end{align}
The first-order formula generalises to higher orders \cite{childsTheoryTrotterError2021};
higher orders cost more per step but scale better in $\varepsilon$.
This work uses the second-order variant.
For $\delta$ small enough, the Trotter error on the ground-state energy obeys
\begin{equation}
    \epstrot = C_{gs}\,\delta^{2},
\label{eq:Cgs}
\end{equation}
with $C_{gs}$ a Hamiltonian-dependent (and therefore system-size-dependent) constant.
Larger $\delta$ reduces the $\mathcal{O}(\varepsilon^{-1}\delta^{-1})$ Trotter-step count,
but keeping $\epstrot$ within target requires $\delta < \sqrt{\varepsilon/C_{gs}}$,
so a tight estimate of $C_{gs}$ is what fixes the optimal step size.
Fully deterministic Trotter also has explicit $L$-dependence,
and $L \sim N^4$ for electronic-structure Hamiltonians with $N$ spin orbitals (qubits).

Randomised product formulas such as qDRIFT and its generalisations \cite{campbellRandomCompilerFast2019, wanRandomizedQuantumAlgorithm2022a} trade the explicit $L$-dependence for $\lambda^2$-scaling at the cost of worse $\varepsilon$-scaling ($\varepsilon^{-2}$ vs.\ $\varepsilon^{-3/2}$ at second order).
They outperform deterministic product formulas when many small-weight terms dominate the Hamiltonian.

The partially randomised method \cite{gunther2025phase} combines both:
high-weight terms enter deterministically,
small-weight terms are sampled.
The Hamiltonian splits as
\begin{align}
\label{eq:hamil_split}
    H = H_D + H_R = \sum_{l=1}^{L_{det}}h_l P_l + \sum_{l = L_{det} + 1}^L h_l \tilde{P_l},
\end{align}
with $L_{det} \ll L$ and $\lambda_R \ll \lambda$;
the total cost takes the form
\begin{align}
    C_{tot} = C_D + C_R = a\frac{L_D}{\varepsilon\delta} + b\frac{\lambda_R^2}{\varepsilon^2},
\end{align}
where $a$ and $b$ are constants set by the phase-estimation protocol.
The Trotter-step constraint still applies,
so large $C_{gs}$ still inflates $C_{tot}$.
The numerical question addressed in this study is how $C_{gs}$ compares against its analytic upper bound in this partially randomised setting.
Because $H_R$ groups many Hamiltonian terms into a single randomised contribution --- unlike standard Trotter formulas --- classical intuition about $C_{gs}$ does not transfer \textit{a priori}.

\subsection*{\label{sec:hamiltonian_prep}Hamiltonian preparation}
The Hamiltonian data used for the simulation represents an embedded fragment of the biologically active ruthenium complex NKP-1339 \cite{Trondl2014, Peti1999} in the binding pocket of the protein GRP78 \cite{Macias2011} that is based on a hybrid model embedding protocol developed in Refs. \cite{Q4Bio-ML,Q4Bio-Embedding}.
To obtain this hybrid model,
two-level QM/QM/MM embedding was employed,
where the large QM region in the initial QM/MM model consisted of the entire NKP-1339 complex \cite{Q4Bio-ML},
which was embedded in the protein environment parametrised with the Amber ff99sb*-ILDN force field \cite{best2009optimised,Lindorff-Larsen2010}.
The technical details of this QM/MM electrostatic-embedding model have been described in detail in Ref. \cite{Q4Bio-ML}.
The electronic structure of this large QM region was described \cite{Q4Bio-ML} with Kohn--Sham density functional theory using Perdew,
Burke,
and Ernzerhof's exchange--correlation functional PBE \cite{Perdew96} with Grimme's D3 dispersion correction \cite{Grimme2010a} and Becke--Johnson damping \cite{Grimme2011} using the def2-SVP basis set \cite{Ahlrich2005}.
The resulting QM/MM-optimised atomic coordinates (Cartesian XYZ,
35~atoms) are provided as \texttt{geom.xyz} in the Zenodo record;
Section~S2 of the Supplementary Material describes the file.

In the second-level quantum-in-quantum embedding,
the Hamiltonian for a smaller QM region (so-called \emph{quantum core} \cite{Q4Bio-Embedding}) embedded in the larger QM region of the QM/MM model
was obtained by Huzinaga-type projection embedding,
performed with the Serenity quantum chemistry program \cite{Serenity2018, Niemeyer2022}) along the lines of what we have previously reported on for a different system in Ref. \cite{Q4Bio-Embedding}
(see Fig.~S5 of the Supplementary Material for a molecular structure of the QM regions in the QM/QM/MM model).
For this work,
we performed active orbital space selection on the set of occupied valence orbitals assigned to the small QM region and on all virtual orbitals of the large QM region using the autoCAS algorithm \cite{Stein2016,Stein2019,Bensberg2024b}.
The autoCAS algorithm selects active orbitals based on an approximate density matrix renormalisation group calculation \cite{White1992} with bond dimension $250$ using the QCMaquis program \cite{Keller2015} (interfaced with the PySCF program \cite{Sun2018} for the calculation of the integrals in the Hartree-Fock molecular orbital basis).
Different active spaces were determined on the basis of the largest single-orbital entropy values obtained from the autoCAS calculation.
The Hamiltonians obtained in this way describe the system with $5$ to $20$ spatial orbitals,
mapped via the canonical Jordan-Wigner transformation to a quantum register of 10--40 qubits,
respectively.
The number of terms of the mapped Hamiltonian ranges from 875 (10 qubits) to 229,140 (40 qubits).
The symmetry-shift technique \cite{loaizaReducingMolecularElectronic2023},
specifically a variant described in \cite{gunther2025phase},
was used to reduce the value of $\lambda$.

To demonstrate that guiding states with high overlap can be efficiently prepared and to generate the reference data,
DMRG calculations were converged for the constructed Hamiltonians with a bond dimension set to $1024$,
which was found to be sufficient to ensure convergence of the DMRG energies for the active spaces considered.
The ordering of the orbitals on the DMRG lattice was determined from the Fiedler vector of the mutual information \cite{barcza2011Fiedler} obtained from the autoCAS calculation \cite{Stein2019, Bensberg2024b}.
The matrix product state (MPS) wavefunctions obtained from these converged DMRG calculations (which will be referred to as the reference MPSs) were taken as a good approximation for the exact ground states for the overlap estimation.
The guiding state preparation strategy involved the reference wavefunction expansion into the Slater determinants basis and taking into consideration only a small number of the determinants with the largest contribution to the expansion.
The most significant determinants and the corresponding coefficients were calculated from the reference MPSs with the sampling-reconstruction of the complete active space (SR-CAS) algorithm \cite{boguslawski2011srcas}.

Finally,
the active Hamiltonians are partitioned into a deterministic and a randomised part.
The optimal split is found by minimising $C_{tot}$ in Eq. \eqref{eq:hamil_split} with respect to $L_D$,
with the constraint that $\epstrot = C_{gs}\delta^2$ must not be larger than $\varepsilon$ (we are using $\varepsilon=0.001$ throughout this work).
The dependency is circular: the optimal partitioning requires $C_{gs}$,
but $C_{gs}$ is only defined once a partitioning is chosen.
A numerical fit of $C_{gs}$ to $\lambda$ from \cite{gunther2025phase} breaks the circularity and fixes the partitioning.
The fit is heuristic but serves as a reasonable proxy,
and the exact value of $C_{gs}$ is then computed for the partitioning it selects.
See Table~S2 of the Supplementary Material for the resulting sizes of the deterministic parts (corresponding to a term order with decreasing magnitude of the coefficients).

\subsection*{\label{sec:data_analysis}Data analysis}
For a general Hamiltonian decomposition $H = \sum_{\alpha} H_{\alpha}$,
an upper bound on the Trotter-error prefactor is \cite{poulinTrotterStepSize2015}
\begin{align}
    C_{gs} \leq 4 \sum_{\alpha=1} \norm{H_{\alpha}} \Bigg(\sum_{\beta: [H_{\alpha}, H_{\beta}]\neq 0} \norm{H_{\beta}}\Bigg)^2,
\end{align}
with $\norm{\cdot}$ the spectral norm.
Specialising to the partially randomised decomposition of Eq.~\eqref{eq:hamil_split},
under the assumption that all deterministic terms fully anticommute with $H_R$,
gives
\begin{align}
    C_{gs} \leq 4 \sum_{l=1}^{L_{det}}\abs{h_l} \Big(\sum_{k:[\tilde{P}_l,\tilde{P}_k]\neq 0}^{L_{det}} \abs{h_k} \Big) + \lambda_R(\lambda-\lambda_R)^2 + (\lambda-\lambda_R)\lambda_R^2.
\end{align}
This bound,
evaluated for the ruthenium-ligand active spaces,
is plotted in Fig.~\ref{fig:trotter-error-scaling}.

$C_{gs}$ is estimated empirically by sampling several step sizes $\delta$ and computing the ground-state energy of the corresponding effective Hamiltonian $\tilde{H}(\delta)$ for each.
For a given $\delta$,
the signal
\begin{align}
    g(t) = \bra{\psi}e^{i\tilde{H}(\delta) t}\ket{\psi},
\end{align}
with $\tilde{H}(\delta)$ as in Eq.~\eqref{eq:define trotter hamiltonian} and partitioning as in Eq.~\eqref{eq:hamil_split},
has its dominant frequency at the effective ground-state energy $\tilde E_0$ ---
the initial state $\ket{\psi}$ is chosen with large ground-state overlap.
RPE \cite{kimmel2015robust} recovers $\tilde E_0$ from an exact simulation of the partially randomised time evolution;
DMRG-computed ground-state energies serve as the exact references $E_0$.
Fewer RPE rounds are required to resolve a larger Trotter error $\epstrot = |E_0 - \tilde E_0|$,
and since $\epstrot = C_{gs}\delta^2$ larger $\delta$ cuts simulation depth.
Fig.~S6 of the Supplementary Material illustrates the extraction procedure.

The signal used by RPE round $m$ is
\begin{align}
    Z_m = g(\delta 2^m) = \bra{\psi}e^{i\delta\tilde{H}(\delta)2^m}\ket{\psi}.
\end{align}
The maximal round $M$ is chosen large enough for $\epstrot$ to converge;
the procedure repeats with a larger $M$ if it does not.
Deterministic terms are ordered lexicographically,
which accelerates their execution via Eq.~\eqref{eq:core}.
The Trotter error depends in principle on the term ordering,
but prior work found no systematic difference between lexicographic and alternative orderings \cite{tranterComparisonBravyiKitaev2018}.

The randomised part follows \cite{gunther2025phase}.
With $\kappa = \delta/(0.2\pi)$,
the number of samples per Trotter stage is $r = \lceil\kappa \lambda_R^2 \delta^2 2^M\rceil$,
and $2r \times 2^m$ terms from $H_R$ are sampled for RPE round $m$ (the extra factor of $2$ reflects the second-order Trotter method).
Randomised terms are not reordered,
since their independence is required for a correct implementation of $e^{-i\delta H_R}$ \cite{gunther2025phase},
which makes their execution slower term-for-term;
for the 20-orbital system, the randomised part is up to $100\times$ slower and becomes the bottleneck.
Above 14 orbitals,
the number of randomised samples is reduced by a factor of up to three to bound the runtime cost.
Lowering the sample count dampens the signal amplitude,
but RPE estimates the complex phase rather than the magnitude,
so a dampened signal still yields a correct energy;
Fig.~S7 of the Supplementary Material illustrates this for a 14-orbital active space.

\subsection*{\label{sec:benches}Benchmarking methodology}

QuEST\,v4,
released shortly before this work's completion,
implements optimisations (i) and (ii) of section \nameref{sec:software_impl} but not the common-suffix grouping of Eq.~\eqref{eq:core};
the comparisons here use v3.
The benchmark compares our implementation with QuEST version 3.7.0
and measures the time needed to perform a group of $L=1$,
$10$,
or $100$ Pauli rotations that share a common suffix $\tilde{Q}$ as in Eq.~\eqref{eq:core}.
The number of qubits ranges from $n=16$ to $n=31$,
and the number of CPU cores is $1$,
$8$ and $128$.
A similar measurement was performed for the GPU-based computation,
with the number of GPU devices: $1$ and $8$.

It should be noted that although we strive to make the computation using our software and QuEST as comparable as possible,
the two versions do not perform exactly the same calculation intrinsically, despite our best effort.
For example,
the implementation of the Pauli rotation $e^{i \varphi \tilde{P}}$,
given by the function \texttt{statevec\_multiRotatePauli()}
as of QuEST version 3.7.0, is less efficient than ours,
even in the most straightforward single-CPU case.
The same benchmark application is linked with four object files,
each containing the compiled and optimised version of the core routine to perform Pauli rotations: using our implementation in the CPU and GPU case,
and using QuEST compiled with MPI or CUDA support,
using \texttt{cmake} compile flags: \texttt{-DDISTRIBUTED=ON} and \texttt{-DGPUACCELERATED=ON} respectively.
Since as of version 3.7.0,
QuEST lacks direct multi-GPU support,
for our benchmark case involving 8 GPUs,
QuEST routines perform the computation locally on individual GPUs,
and the MPI data transfer is handled by the benchmark application,
mimicking QuEST's approach for the CPU-based computation.
We perform the same computation for our implementation and for QuEST,
measuring the duration needed to apply the same set of Pauli rotations 10 times on the array of amplitudes
and dividing the total measured duration by 10 to obtain the average over memory cache effects.
Because the exact values of the quantum amplitudes are irrelevant for timing the execution,
we initialise the simulation state with random values.
The measurements were performed on Gefion's single NVIDIA DGX H100 server
with 2 Intel Xeon Platinum 8480C CPUs and 8 NVIDIA H100 GPUs.

\newpage
\section*{\label{sec:data_avail}Data availability}

The Hamiltonian data (ruthenium complex NKP-1339),
the optimised atomic coordinates (Cartesian XYZ format,
35~atoms),
simulation results,
and benchmark data that support the findings of this study are
openly available in the Zenodo repository:
\url{https://doi.org/10.5281/zenodo.19690600}~\cite{miller2026phase2data}.

\section*{\label{sec:code_avail}Code availability}

The source code of the simulator software \texttt{phase2} is available under the
BSD-3-Clause license.
Version 0.13.1 was used to perform the experiments reported here;
the archived release is deposited in the Zenodo repository~\cite{miller2025phase2code}.
Version 1.0 adds a Python wrapper and usability improvements.

\section*{\label{sec:acknowledgements}Acknowledgements}
The authors would like to thank Moritz Bensberg for his valuable assistance with the preparation of the Hamiltonian input data,
and Anders Krogh for help with securing computational resources.
Moreover,
we are grateful to Olivier Byrde and the Euler team at ETH Z\"urich for their help in tailoring the accessibility of the Euler cluster to the needs of this work.
We also gratefully acknowledge the support of the Danish Centre for AI Innovation (DCAI),
which hosts the Gefion supercomputer,
where part of the computations were done in the frame of the pilot project ``Large-scale distributed simulation of quantum algorithms for quantifying molecular recognition processes''.

\section*{\label{sec:funding}Funding}
Funding from the Research Project ``Molecular Recognition from Quantum Computing'' is acknowledged.
Work on ``Molecular Recognition from Quantum Computing'' was supported by Wellcome Leap as part of the Quantum for Bio (Q4Bio) Program.
M.C.,
J.G.,
F.W.
and M.M.
also acknowledge financial support from the Novo Nordisk Foundation (grant no.\ NNF20OC0059939 ``Quantum for Life'').
M.S.T.,
G.C.S.
acknowledge support from the Novo Nordisk Foundation,
(grant no.\ NNF22SA0081175,
NNF Quantum Computing Programme and grant no.\ NNF20OC0060019,
Solid-Q).
M.M.,
M.C.
acknowledge support from the novoSTAR Programme by NovoNordisk A/S.

\section*{\label{sec:contribution}Author contributions}
M.M.
designed and wrote the simulation software.
M.C.,
J.G.,
M.M.
and F.W.
conceptualised the research project.
J.G.,
M.M.
and M.S.T.
conducted the numerical experiment and analyzed data.
M.E.
prepared Hamiltonian data.
M.C.,
M.R.
and G.C.S.
acquired computational resources.
M.E.,
J.G.,
M.M.
and F.W.
wrote the first version of the manuscript.
All authors reviewed the manuscript.

\section*{\label{sec:Competing}Competing interests}
The authors declare no competing interests.

\bibliographystyle{naturemag}
\bibliography{refs}

\newpage
\setcounter{figure}{0}
\setcounter{table}{0}
\setcounter{equation}{0}
\renewcommand{\thefigure}{S\arabic{figure}}
\renewcommand{\thetable}{S\arabic{table}}
\renewcommand{\theequation}{S\arabic{equation}}

\begin{center}
{\Large\bfseries Supplementary Material}
\end{center}

\section*{\label{sec:supp_source_data}S1.\ Source data for main-text figures}

Source data for the main-text figures are deposited as CSV files,
one per figure,
in the Zenodo record
[DOI \href{https://doi.org/10.5281/zenodo.19690600}{10.5281/zenodo.19690600}],
alongside the Hamiltonian,
simulation and benchmark datasets.

\begin{itemize}
\item \texttt{q4bio\_simul-20250407.csv} — Fig.~1 (weak scaling).
  One row per SLURM-job checkpoint;
  columns include \texttt{job\_id, backend, n\_qb, n\_cpus, steps, t\_run}.
\item \texttt{phase2-benchmark\_quest.csv} — Fig.~2 (performance comparison).
  One row per benchmark cell;
  columns include \texttt{arch, backend, n\_qb, n\_tasks, n\_paulirot, t\_ms}.
\item \texttt{ruth-group-size-histogram-n40.csv} — Fig.~3 (common-suffix histogram).
  One row per $(k,\,\text{group index})$ pair;
  column \texttt{group\_size} records the number of Hamiltonian terms in that group.
\item \texttt{trotter-error-fit.csv} — Fig.~4 (Trotter-error prefactor scaling).
  One row per active space;
  columns include \texttt{num\_orbitals, lambda, C\_gs\_fit, a\_fit, C\_gs\_bound}.
\end{itemize}

\section*{\label{sec:supp_coordinates}S2.\ Atomic coordinates}

The optimised atomic coordinates of the ruthenium complex
NKP-1339 used in the electronic-structure calculations are
deposited as the file \texttt{geom.xyz} (XYZ format,
35~atoms) in the Zenodo record at
\url{https://doi.org/10.5281/zenodo.19690600}.
The geometry corresponds to the QM/MM-optimised structure of the
NKP-1339 complex described in Ref.~\cite{Q4Bio-ML} and in the
Hamiltonian-preparation subsection of the main text.

\section*{\label{sec:supp_figures}S3.\ Supplementary figures and tables}

\subsection*{Strong scaling}

\begin{figure}[h!]
\centering
\includegraphics[width=0.75\linewidth]{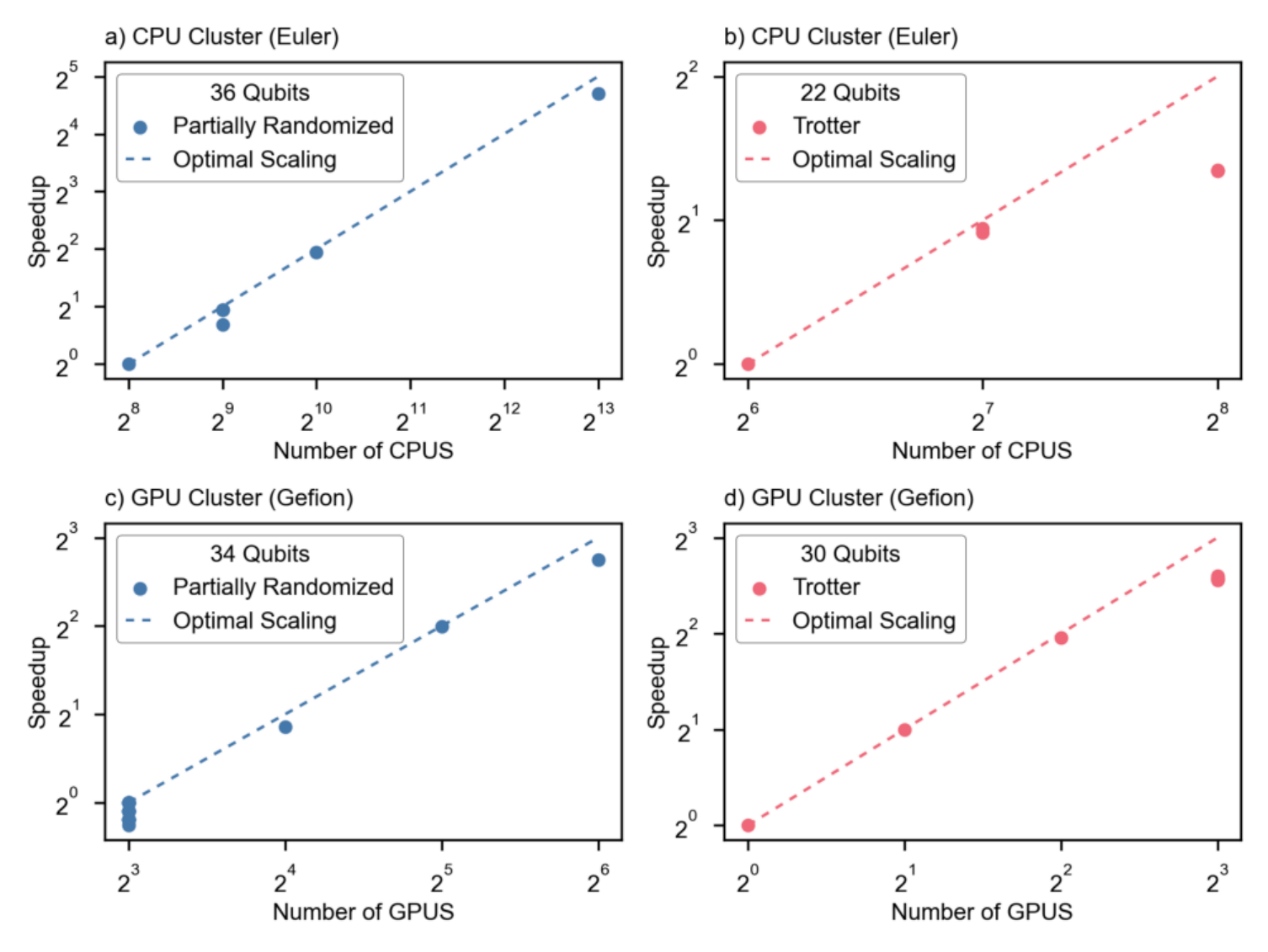}
\caption{Strong scaling.
Speedup versus number of CPU cores (\textbf{a},\,\textbf{b}; Euler) or GPU devices (\textbf{c},\,\textbf{d}; Gefion) for a fixed problem size.
Blue: partially randomised algorithm; red: deterministic Trotter formula.
Dashed lines: ideal linear scaling.}
\label{fig:supp-strong-scaling}
\end{figure}

Fig.~\ref{fig:supp-strong-scaling} reports the strong scaling of
the simulation algorithm:
for a fixed problem size,
determined by the number of qubits and the number of Pauli
rotations,
the speedup of the total simulation time is plotted as a function
of the amount of parallel computational resources.

\subsection*{Single-CPU comparison with Qiskit Aer}

The memory-bound regime of full-state vector simulation can be
observed directly by comparing \texttt{phase2} against two
established simulator backends on a single multicore CPU.
Differences in threading strategy show at small system sizes;
once the simulation state outgrows the CPU cache,
memory access dominates and all three backends run at comparable
speed.
In distributed simulations,
\texttt{phase2} amortises the memory-access cost over common-suffix groups of Pauli rotations (Software implementation and Performance sections),
whereas the three backends compared here apply elementary operations independently.

\begin{figure}[h!]
\centering
\includegraphics[width=0.5\textwidth]{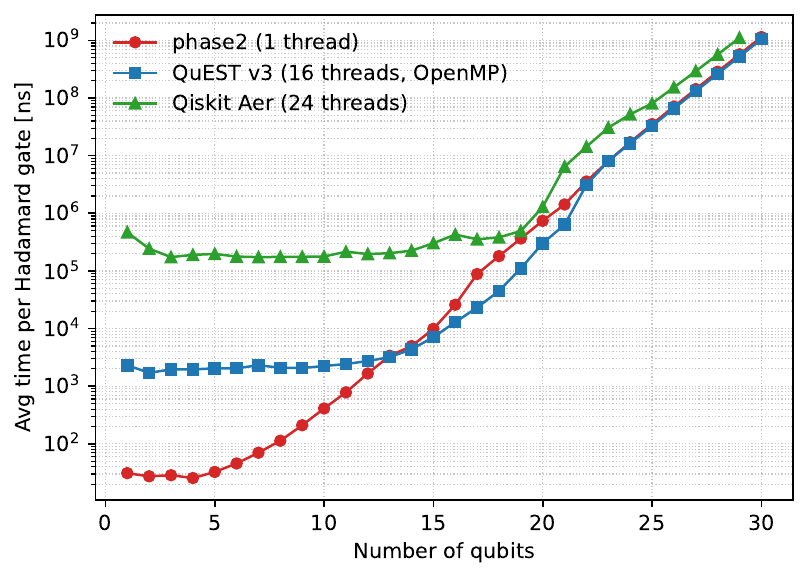}
\caption{Single-CPU benchmark: average time (ns) per Hadamard gate versus number of qubits.
\texttt{phase2} (1 thread),
QuEST\,v3 (16 OpenMP threads),
Qiskit Aer (24 threads);
Intel Core i9-12900,
5.00\,GHz.
All three converge at large $n$,
confirming that performance is memory-bound.}
\label{fig:supp-single-cpu-benchmark}
\end{figure}

\subsection*{Common-suffix group size scaling for the ruthenium-ligand Hamiltonian}

\begin{figure}[h!]
\centering
\includegraphics[width=\linewidth]{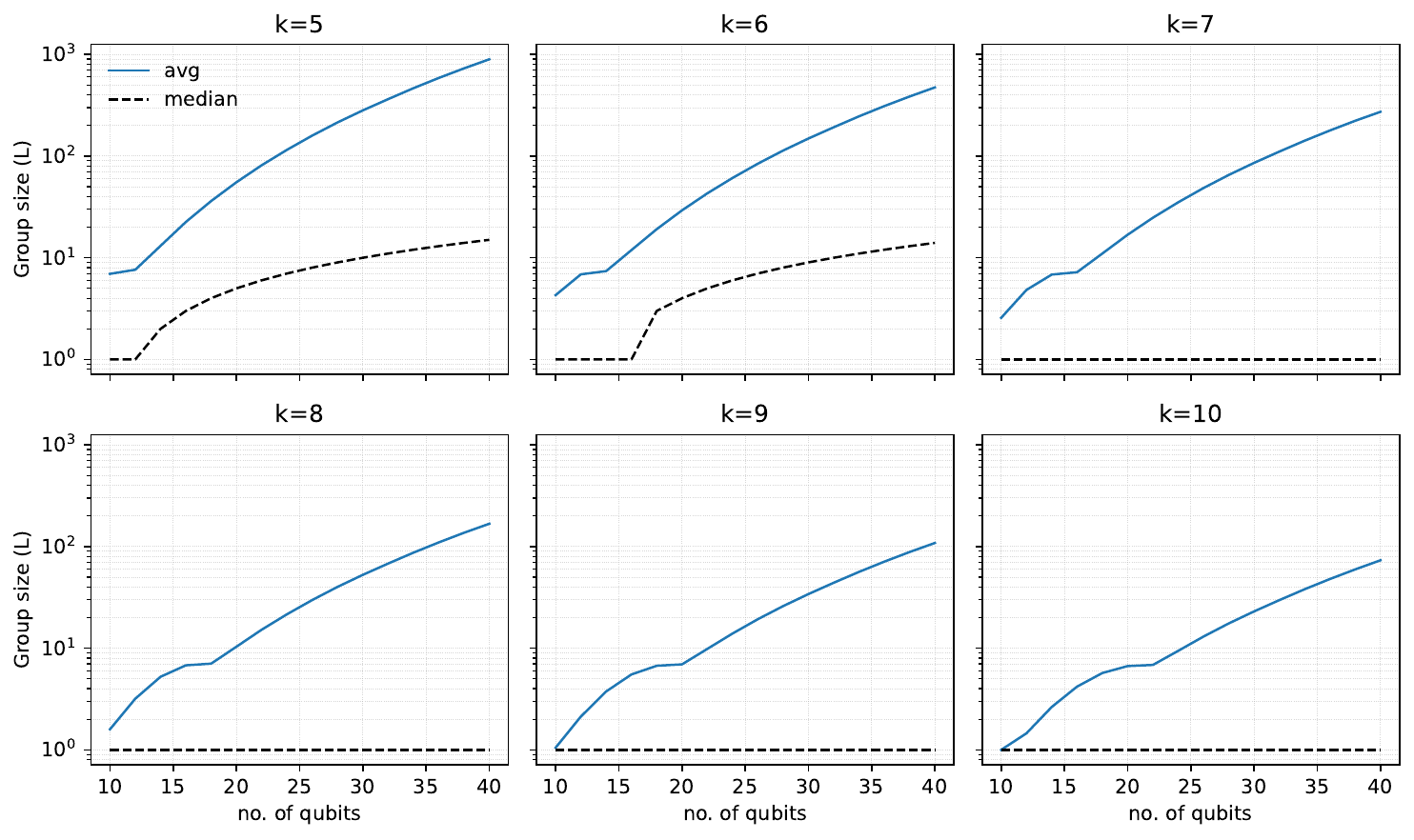}
\caption{Common-suffix group size versus number of qubits $n$ for suffix lengths $k = 5$--$10$.
Solid lines: average; dashed lines: median.}
\label{fig:supp-ruth-scaling}
\end{figure}

Fig.~\ref{fig:supp-ruth-scaling} shows how the average and median
common-suffix group size scale with the number of qubits for the
ruthenium-ligand Hamiltonian,
across suffix lengths $k = 5, 6, \ldots, 10$.
For longer suffices ($k \geq 7$),
the median remains close to~1 while the average grows
exponentially,
reflecting a heavy-tailed distribution of group sizes
on which the distributed simulation depends for its
speedup.

\subsection*{Largest common-suffix groups at $n = 40$, $k = 9$}

Table~\ref{tab:supp-ruth-top-groups} lists the 30 largest
common-suffix groups for suffix length $k = 9$
(corresponding to the $2^{9} = 512$-GPU partitioning used on
Gefion).
The dominant group ($\tilde{Q} = I^{\otimes 9}$,
rank~1) alone accounts for 37.9\,\% of all Hamiltonian terms;
the top~19 groups collectively cover 76.4\,\% of the
229\,140 terms.

\begin{table}[h!]
    \centering
    \small
    \begin{tabular}{|r|l|r|r|}

    \hline
        Rank & Suffix $\tilde{Q}$ & Group size $L$ & Cumul.\ fraction \\ \hline
         1 & \texttt{IIIIIIIII} & 86\,886 & 0.3792 \\
         2 & \texttt{ZZZZZZXII} &  4\,963 & 0.4008 \\
         3 & \texttt{ZZZZZZYII} &  4\,963 & 0.4225 \\
         4 & \texttt{ZZZZYIIII} &  4\,961 & 0.4442 \\
         5 & \texttt{ZZZZXIIII} &  4\,961 & 0.4658 \\
         6 & \texttt{ZZYIIIIII} &  4\,914 & 0.4872 \\
         7 & \texttt{ZZXIIIIII} &  4\,914 & 0.5087 \\
         8 & \texttt{YIIIIIIII} &  4\,893 & 0.5300 \\
         9 & \texttt{XIIIIIIII} &  4\,893 & 0.5514 \\
        10 & \texttt{ZXIIIIIII} &  4\,892 & 0.5728 \\
        11 & \texttt{ZYIIIIIII} &  4\,892 & 0.5941 \\
        12 & \texttt{ZZZZZYIII} &  4\,886 & 0.6154 \\
        13 & \texttt{ZZZZZXIII} &  4\,886 & 0.6367 \\
        14 & \texttt{ZZZYIIIII} &  4\,872 & 0.6580 \\
        15 & \texttt{ZZZXIIIII} &  4\,872 & 0.6793 \\
        16 & \texttt{ZZZZZZZZY} &  4\,871 & 0.7005 \\
        17 & \texttt{ZZZZZZZZX} &  4\,871 & 0.7218 \\
        18 & \texttt{ZZZZZZZXI} &  4\,856 & 0.7430 \\
        19 & \texttt{ZZZZZZZYI} &  4\,856 & 0.7642 \\
        20 & \texttt{IYZZZZYII} &    522 & 0.7664 \\
        21 & \texttt{IZIIIIIII} &    522 & 0.7687 \\
        22 & \texttt{IXZZZZXII} &    522 & 0.7710 \\
        23 & \texttt{XZZXIIIII} &    522 & 0.7733 \\
        24 & \texttt{IIIIIIIZI} &    522 & 0.7756 \\
        25 & \texttt{IIIIIIIXX} &    522 & 0.7778 \\
        26 & \texttt{IIIIIIIYY} &    522 & 0.7801 \\
        27 & \texttt{IIIIIIIIZ} &    522 & 0.7824 \\
        28 & \texttt{IIIIZIIII} &    522 & 0.7847 \\
        29 & \texttt{IIIXZZXII} &    522 & 0.7870 \\
        30 & \texttt{IIIYZZYII} &    522 & 0.7892 \\ \hline
    \end{tabular}
    \caption{Thirty largest common-suffix groups for the
    ruthenium-ligand Hamiltonian at $n = 40$ qubits and suffix
    length $k = 9$ (corresponding to $2^{9} = 512$ GPUs).
    The suffix $\tilde{Q}$ is shown in $I, X, Y, Z$ notation.
    The cumulative fraction is the running share of the
    229\,140 total Hamiltonian terms covered by the top-ranked
    groups.}
    \label{tab:supp-ruth-top-groups}
\end{table}

\subsection*{Signal amplitude for the 40-qubit partially randomized simulation}

\begin{figure}[h!]
\centering
\includegraphics[width=0.5\linewidth]{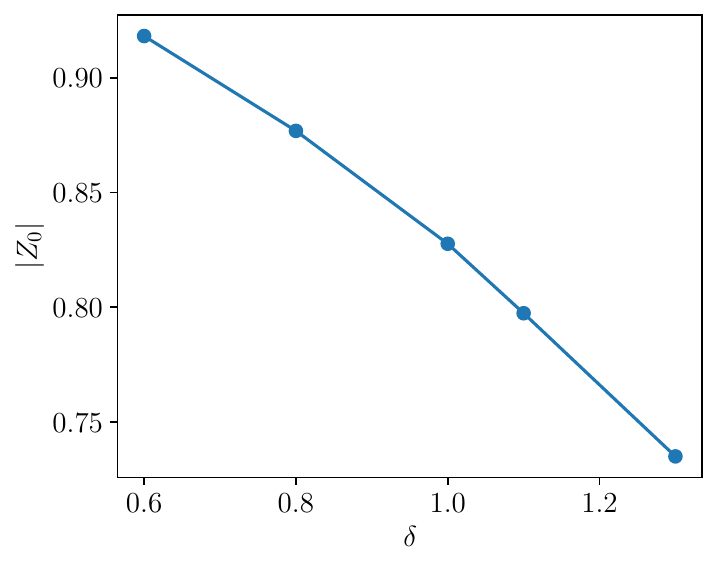}
\caption{Signal amplitude $Z_0 = \bra{\psi}e^{i\delta\tilde{H}(\delta)}\ket{\psi}$ versus step size $\delta$ for the 40-qubit (20-orbital) simulation;
$1.26 \times 10^{4}$ Pauli rotations per data point.}
\label{fig:supp-cas20-amplitude}
\end{figure}

Fig.~\ref{fig:supp-cas20-amplitude} reports the amplitude of the
signal
$Z_0 = \bra{\psi}e^{i\delta\tilde{H}(\delta)}\ket{\psi}$
for the 40-qubit simulation of the 20-orbital active space,
performed on 64 NVIDIA DGX H100 nodes with 512 H100 GPUs sharing
the simulation state.
The figure is cited from the main-text Simulation subsection as
evidence that the largest analysed case produces the expected
non-vanishing amplitude,
validating correctness of the distributed implementation in the
regime where an incorrect implementation of
$e^{i\delta\tilde{H}(\delta)}$ would,
with high probability,
collapse the amplitude to near zero.

\subsection*{Molecular structure of the QM regions}

\begin{figure}[h!]
    \centering
    \includegraphics[width=0.5\linewidth]{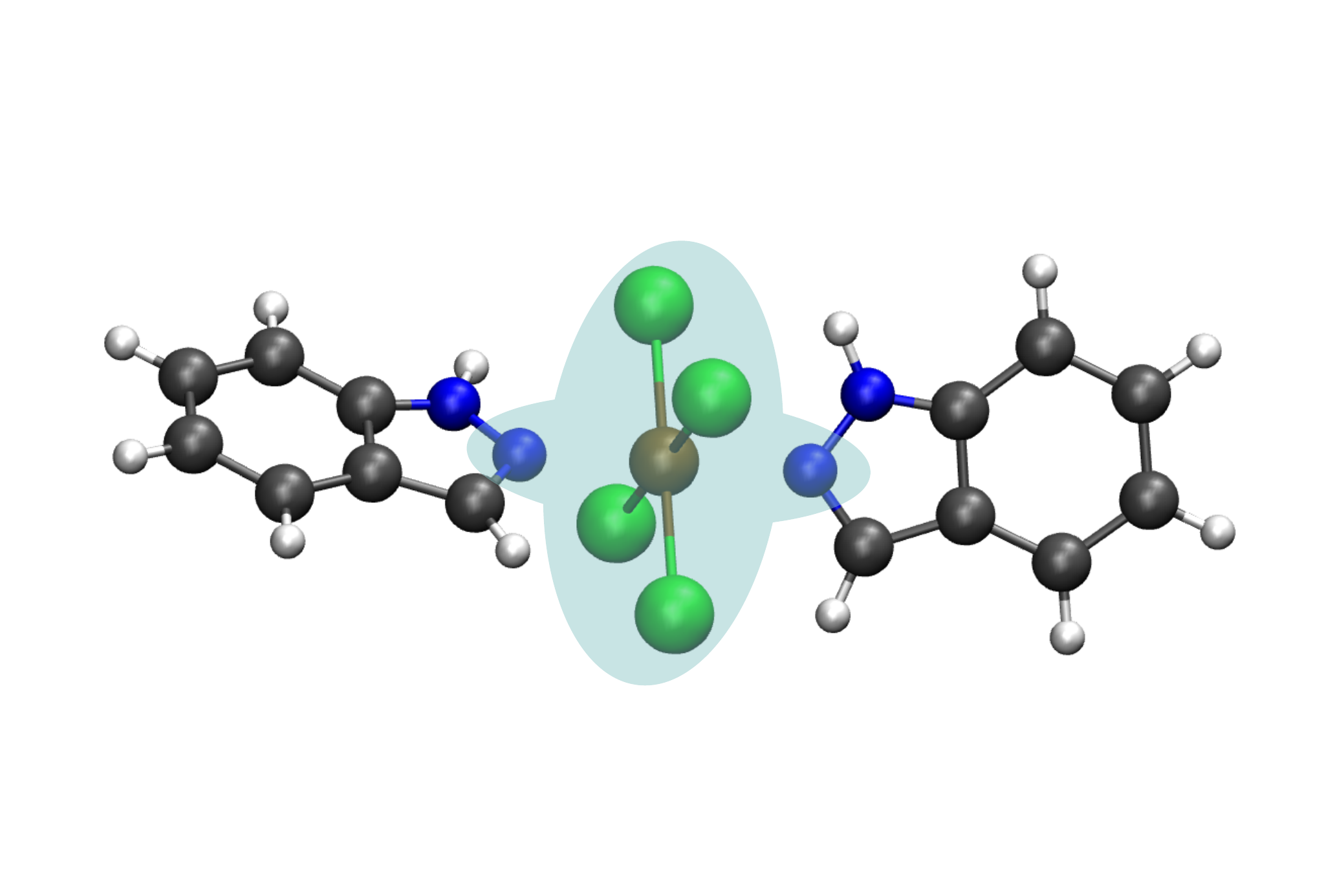}
    \caption{Molecular structure of NKP-1339 from the QM/MM model of Ref.~\cite{Q4Bio-ML}.
    The small QM region used for Hamiltonian construction is shaded blue.
    Atom colours: C black,
    H white,
    N blue,
    Cl green,
    Ru (centre) brown.}
    \label{fig:supp-ru-complex}
\end{figure}

Fig.~\ref{fig:supp-ru-complex} shows the molecular structure of
the ruthenium complex NKP-1339 used throughout the main-text
numerical experiment,
taken from the QM/MM hybrid model described in
Ref.~\cite{Q4Bio-ML} and cross-referenced from the
Hamiltonian-preparation subsection of the main text.
The small QM region that enters the Hamiltonian construction is
highlighted in transparent blue.

\subsection*{Trotter-error analysis walk-through at CAS\_08}

\begin{figure}[h!]
    \centering
    \includegraphics[width=0.9\textwidth]{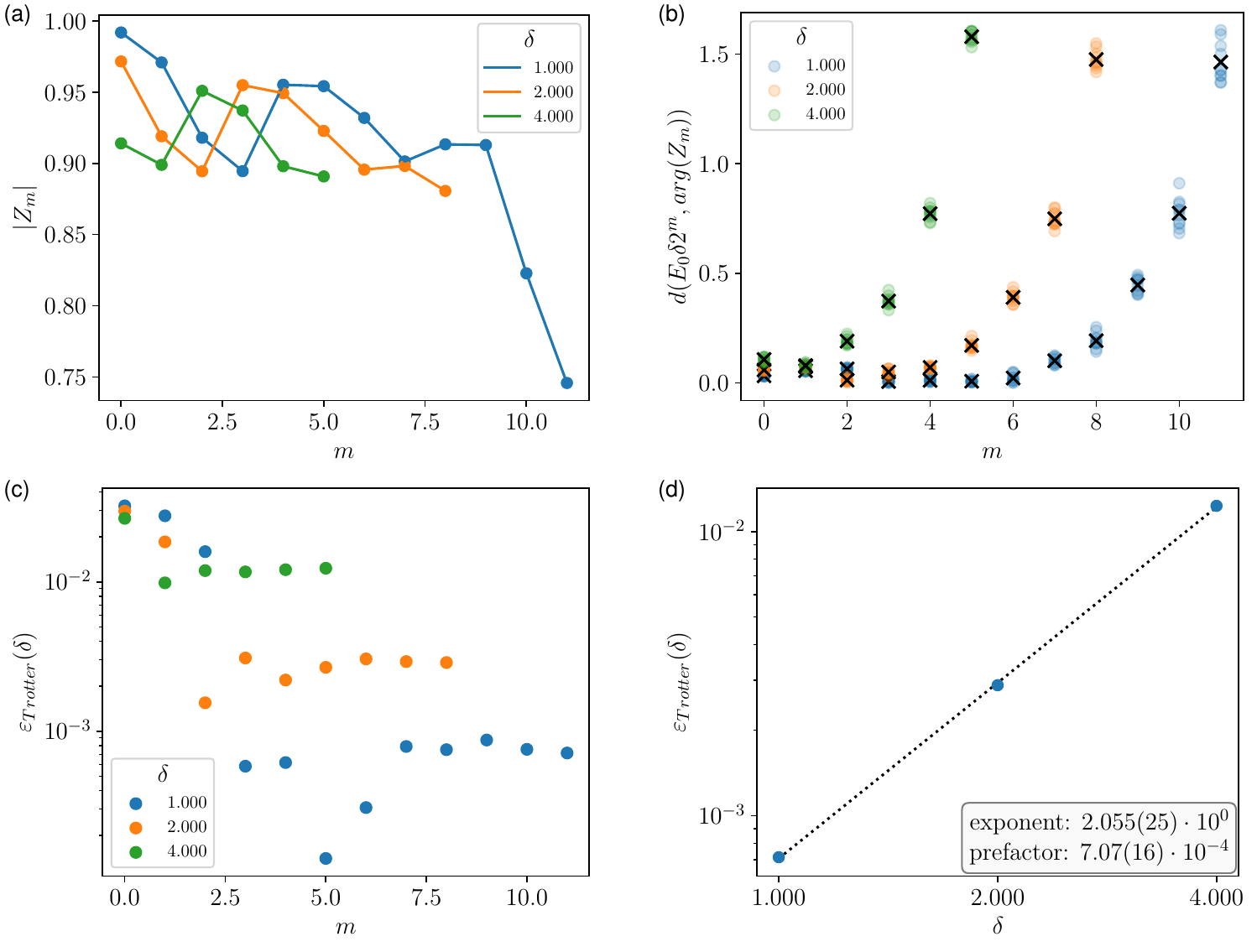}
    \caption{Extraction of the Trotter-error prefactor $C_{gs}$ for CAS\_08.
    \textbf{a}~Signal amplitude $\abs{Z_m}$ versus RPE round $m$ at step size $\delta$.
    \textbf{b}~Phase difference between exact and partially randomised time evolution (circles: individual runs; crosses: averages).
    \textbf{c}~Convergence of $\epstrot$ during RPE.
    \textbf{d}~Power-law fit of $\epstrot(\delta)$ to $\delta$.}
    \label{fig:supp-cas8-trottererror}
\end{figure}

Fig.~\ref{fig:supp-cas8-trottererror} walks through the
numerical procedure used to extract the Trotter-error prefactor
$C_{gs}$ from the simulated time signals $Z_m$ on the 8-orbital
active space (CAS\_08).
It is cited from the Data-analysis subsection of the main text
as a concrete illustration of how the step-size fit that yields
$C_{gs}$ is obtained in practice,
and how the RPE algorithm converges for this representative
system size.

\subsection*{Hamiltonian partitioning and Trotter-error fit parameters}

Table~\ref{tab:supp-trotter_results} lists,
for each active space of the ruthenium-ligand Hamiltonian,
the total number of Pauli terms,
the size of the deterministic partition $L_D$,
the one-norm $\lambda$ and the ratio $\lambda_R / \lambda$ of the
randomized remainder,
the number of RPE rounds $m$ used,
the fitted Trotter-error prefactor $C_{gs}$ and exponent $a$,
the analytic $C_{gs}$ upper bound,
and the largest simulated circuit in Pauli rotations.
The entries are the source data for the Trotter-error scaling
figure in the main text and for the optimal-partitioning
discussion in the Hamiltonian-preparation subsection.

\begin{table}[h!]
    \centering
    \small
    \begin{tabular}{|c|c|c|c|c|c|c|c|c|c|}

    \hline
        no. orb. & no. tot. & no. det. & $\lambda$ & $\lambda_R/\lambda$ & RPE m & $C_{gs}$ fit & $a$ fit & $C_{gs}$ bound & max circ.\\ \hline
        5 & 875 & 155 & 2.6 & 0.079 & 18 & $5.90\cdot 10^{-4}$ & 2.01 & $8.80 \cdot 10^0$ & $1.25\cdot10^9$\\
        6 & 1800 & 170 & 3.1 & 0.11 & 15 & $3.53 \cdot 10^{-4}$ & 2.02 & $1.53 \cdot 10^1$ & $2.05\cdot 10^8$ \\
        7 & 3355 & 277 & 4.6 & 0.12 & 13 & $2.95\cdot 10^{-4}$& 1.98 & $4.91 \cdot 10^1$ & $1.75\cdot 10^7$\\
        8 & 5758 & 362 & 6.4 & 0.15 & 12 & $7.07\cdot 10^{-4}$& 2.06 & $1.51\cdot 10^2$ & $1.46\cdot 10^7$ \\
        9 & 9249 & 429 & 7.1 & 0.18 & 12 & $8.51\cdot 10^{-4}$ & 2.06 & $2.23 \cdot 10^2$ & $4.60\cdot 10^7$ \\
        10 & 14156 & 574 & 8.2 & 0.20 & 14 & $1.25\cdot 10^{-3}$ & 2.14 & $3.74 \cdot 10^2$ & $3.10\cdot 10^8$\\
        11 & 20813 & 895 & 10.8 & 0.21 & 13 & $6.12\cdot 10^{-4}$ & 2.40 & $8.82 \cdot 10^2$ & $1.01\cdot 10^9$ \\
        12 & 29512 & 970 & 12.4 & 0.22 & 11 & $6.10\cdot 10^{-4}$ & 2.09 & $1.35 \cdot 10^3$ & $5.10\cdot 10^7$\\
        13 & 40755 & 837 & 16.2 & 0.23 & 10 & $9.39 \cdot10^{-4}$ & 2.27 & $3.04 \cdot 10^3$ & $2.04\cdot 10^7$\\
        14 & 54876 & 872 & 19.8 & 0.23 & 8 & $1.60\cdot 10^{-3}$ & 1.90 & $5.64 \cdot 10^3$ & $4.90\cdot 10^6$\\
        15 & 72071 & 1053 & 23.4 & 0.24 & 9 & $7.34\cdot 10^{-3}$ & 2.27 & $9.48 \cdot 10^3$ & $6.56\cdot 10^6$ \\
        16 & 92968 & 1322 & 27.0 & 0.24 & 8 & $1.28\cdot 10^{-2}$ & 2.24 & $1.45 \cdot 10^4$ & $2.83\cdot 10^6$ \\
        17 & 119087 & 1873 & 31.0 & 0.27 & 6 & $2.84 \cdot 10^{-1}$ * & - & $2.36 \cdot 10^4$ & $1.37\cdot 10^6$ \\
        18 & 150088 & 2396 & 36.0 & 0.28 & 4 &  $5.07\cdot 10^{-1}$ * & - & $3.82 \cdot 10^4$ & $2.19\cdot 10^5$\\
        19 & 186561 & 2853 & 41.5 & 0.29 & 3 & - & - & $5.98\cdot 10^4$ & $9.56\cdot 10^4$\\
        20 & 229140 & 3284 & 48.9 & 0.29 & 1 & $8.17\cdot 10^{-1}$ * & - & $9.83 \cdot 10^4$ & $1.26\cdot 10^4$ \\\hline
\end{tabular}
    \caption{Total and deterministic terms of the Hamiltonians,
    fitting parameters of the Trotter error ($\epstrot = C_{gs}\delta^a$),
    and the maximum circuit simulated,
    in terms of number of Pauli rotations.
    The asterisk (*) indicates the lack of convergence,
    and hence no fit for $a$,
    due to insufficient numerical data.}
    \label{tab:supp-trotter_results}
\end{table}

\subsection*{Sensitivity of the partially randomized method to the number of random samples}

\begin{figure}[h!]
\centering
\begin{subfigure}[b]{0.45\textwidth}
\includegraphics[width=\textwidth]{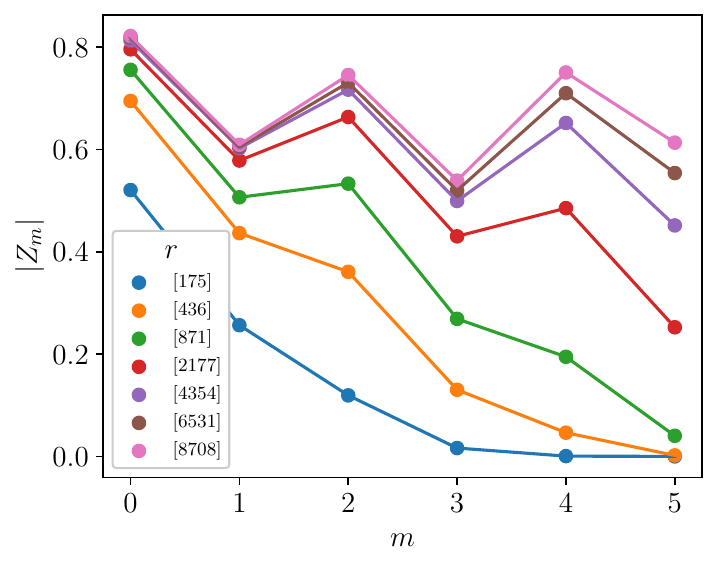}
\caption{Signal amplitude $\abs{Z_m}$ versus RPE round $m$ for varying $r$.}
\end{subfigure}\hfill
\begin{subfigure}[b]{0.45\textwidth}
\includegraphics[width=\textwidth]{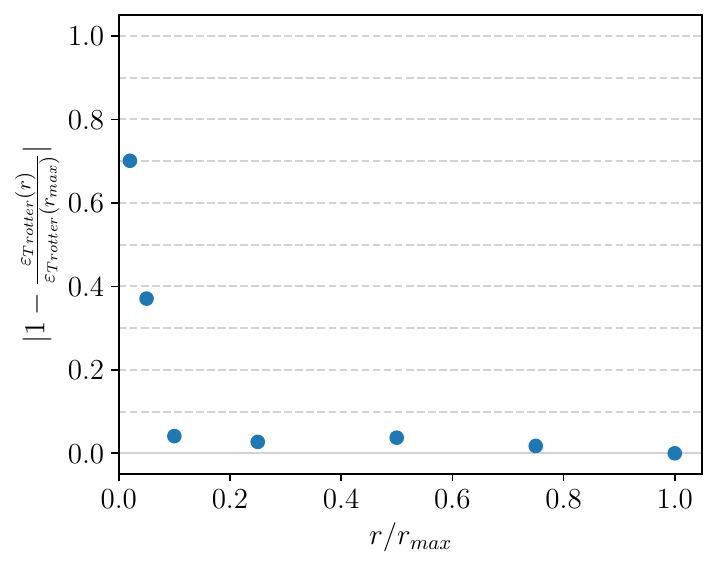}
\caption{Relative error in $\epstrot$ versus $r$,
normalised to $r_{\max}$.}
\end{subfigure}
\caption{Sensitivity to the number of random samples $r$ for the 14-orbital active space at $\delta = 4$.}
\label{fig:supp-cas14-numsamples}
\end{figure}

Fig.~\ref{fig:supp-cas14-numsamples} documents how reducing the
number of random samples $r$ in the partially randomized Trotter
product formula (used throughout the main-text numerical
experiment) trades off against accuracy of the extracted
Trotter-error prefactor.
The data in the main text claim that RPE estimates the complex
phase,
not the magnitude,
of the signal,
so a signal with a dampened amplitude still yields a correct
energy estimate;
this figure substantiates that claim on the 14-orbital active
space.

\section*{S4.\ Supplementary Methods}

\subsection*{Distributed Pauli-rotation derivations}

This section provides the full derivations summarised in the
Mathematical model subsection of the main text.

\subsubsection*{State decomposition into lower and upper qubits}

Dividing the Hilbert space into $n-m$ ``lower'' and $m$ ``upper'' qubits:
$\mathcal{H} = (\mathbb{C}^{2})^{\otimes n-m} \otimes  (\mathbb{C}^{2})^{\otimes m}$,
by writing a computational basis state $|i\rangle$ as:
$|i\rangle = |i_{0} i_{1} \ldots i_{n-m} \rangle \, |i_{n-m+1} \ldots i_{n-1}\rangle$,
where $(i_{n-1} \ldots i_{1} i_{0})_{2}$ is the binary representation of $i$,
$i = 0, 1, \ldots 2^{n} - 1$,
we can write the quantum state $|\psi\rangle$ in the form:
\begin{equation}
|\psi\rangle = \sum \limits_{k=0}^{M-1} \sum \limits_{i=0}^{N-1} A_{k}(i) \, |i\rangle |k\rangle= \
\sum \limits_{k=0}^{M-1} \sum \limits_{i=0}^{N-1} a_{i + k \cdot 2^{n-m}} |i\rangle |k\rangle.
\end{equation}

\subsubsection*{Local operations on lower qubits}

A unitary operator $U \otimes I_{m}$,
acting as an identity on the upper qubits,
can be implemented by the same matrix operation performed by all MPI tasks in parallel on their respective partitions $A_{k}$.
If $U_{ij}$ are matrix elements of the unitary $U$ in the computational basis,
$U_{ij} = \langle i | U | j \rangle$,
$i,j = 0,1, \ldots, N-1$,
we have
\begin{equation}
U \otimes I_{m} |\psi\rangle = \sum \limits_{k=0}^{M-1} \sum \limits_{j=0}^{N-1} A_{k}(j) \, U |j\rangle |k\rangle=
 \sum \limits_{k=0}^{M-1} \sum \limits_{i, j=0}^{N-1} U_{ij} A_{k}(j) \, |i\rangle |k\rangle.
\end{equation}
Hence,
the action of $ U \otimes I_{m}$ on $|\psi\rangle$ amounts to a local transformation
$[A_{k}(i)] \mapsto [ \sum_{j} U_{ij} A_{k}(j) ]$
of the partitions of the array by each MPI task in parallel.

\subsubsection*{Non-local operations on upper qubits}

To implement a unitary $I_{n-m} \otimes U^{\prime}$,
the tasks need to exchange their partitions with each other.
In a special case where $U^{\prime} = \tilde{P} =  P_{1} \otimes P_{2} \otimes \ldots \otimes P_{m}$,
$P_{k} \in \{ I, \sigma_{x}, \sigma_{y}, \sigma_{z} \}$,
$k = 0,1, \ldots, m$,
because $\tilde{P}^{2} = I_{m}$ and $\tilde{P}^{\dagger} = \tilde{P}$,
the space $(\mathbb{C}^{2})^{\otimes m}$ splits into $2^{m-1}$ $2$-dimensional orthogonal subspaces,
invariant under the action of $\tilde{P}$.
To prove this statement,
let us note first that it is clearly true,
if $\tilde{P}$ is a diagonal operator in the computational basis,
i.e. when all $P_{k}$ are either $I$ or $\sigma_{z}$.
Furthermore,
it is easy to see that if the statement holds true for two Pauli strings $\tilde{P}_{1}$ and $\tilde{P}_{2}$
such that $\tilde{P}_{1} \tilde{P}_{2} = \tilde{P}_{2} \tilde{P}_{1}$,
it also holds for $\tilde{P}_{1} \tilde{P}_{2}$,
as Pauli strings are Hermitian operators.
Hence,
we can reduce the proof to the special case when all $\tilde{P}_{k}$ are equal to the identity operator
except for  exactly one of them,
e.g. $\tilde{P} = \sigma_x \otimes I \otimes \ldots \otimes I$.
In that case,
the statement can be verified explicitly by writing out the matrix elements of $\tilde{P}$ in the computational basis.
By this reasoning,
to achieve the action of $I_{n-m} \otimes \tilde{P}$ on the state $|\psi\rangle$,
the task $k$ needs to obtain only the amplitudes from the task $k^{\prime}$ such that
$\tilde{P} |k\rangle = \omega_{k} |k^{\prime}\rangle$,
where $\omega_{k} \in \{ 1, i, -1, -i \}$,
and \emph{vice versa}.
Note that in the case where $\tilde{P}$ is a diagonal operator,
it might as well happen that $k^{\prime} = k$.
In any case,
the map $k \mapsto k^{\prime}$ is a bijection for every Pauli string,
and hence the action of $I_{n-m} \otimes \tilde{P}$ reduces to a pairwise exchange of partitions $A_{k}$ and $A_{k^{\prime}}$ for $k = 0, 1, \ldots, 2^{m} - 1$.
More precisely,
\begin{equation}
I_{m-n} \otimes \tilde{P} |\psi\rangle = \sum \limits_{k=0}^{M-1} \sum \limits_{i=0}^{N-1} \omega_{k} A_{k}(i) \, |i\rangle |k^{\prime}\rangle =
\sum \limits_{k=0}^{M-1} \sum \limits_{i=0}^{N-1} \overline{\omega_{k}} A_{k^\prime}(i) \, |i\rangle |k \rangle,
\end{equation}
where the last step follows from swapping the indices $k$ and $k^{\prime}$,
since $\tilde{P}$ is a bijection,
and from the fact that $\omega_{k^{\prime}} = \omega_{k}^{-1} = \overline{\omega_{k}}$,
the complex conjugation,
since $\tilde{P}^{2} = I$.
The task $k$ can store the amplitudes $A_{k^{\prime}}$,
multiplied by the factor $\overline{\omega_{k}}$,
in their buffer $B_{k}$.
See also \cite{jones2023distributed} for an alternative derivation of this communication scheme.

\subsubsection*{Proof of the common-suffix grouping identity}

A further enhancement is to observe that the network traffic between MPI tasks can be reduced to a single buffer exchange in the case where a sequence of Pauli rotations share the same Pauli string on upper qubits:
\begin{equation*}
\prod_{l=1}^{L} e^{i \varphi_l \tilde{P}_l \otimes \tilde{Q}} =  \prod_{l=1}^{L} e^{i \varphi_l \tilde{P}_l} \otimes
\frac{I + \tilde{Q}}{2} +
\prod_{l=1}^{L} e^{-i \varphi_l \tilde{P}_l} \otimes
\frac{I - \tilde{Q}}{2}.
\end{equation*}
where $\tilde{P}_l = P_{l,1} \otimes \ldots \otimes P_{l, n-m}$ are Pauli strings that act on the lower qubits and $\tilde{Q} = Q_{1} \otimes Q_{2} \otimes \ldots \otimes Q_{m}$ is a Pauli string on the upper qubits,
which is constant for $l = 1, 2, \ldots, L$.
This fact can be proved by induction over $L$.
As a consequence of the properties of Pauli strings:
$\tilde{P} = \tilde{P}^{\dagger}$,
$\tilde{P}^2 = I$,
$\tilde{Q} = \tilde{Q}^{\dagger}$,
$\tilde{Q}^{2} = I$,
from the definition of the matrix exponent we have
\begin{multline}
e^{i \varphi \tilde{P} \otimes \tilde{Q}} =  \cos \varphi I \otimes I + i \sin \varphi \tilde{P} \otimes \tilde{Q} = \\
= \left( \cos \varphi I + i \sin \varphi \tilde{P} \right) \otimes \frac{I + \tilde{Q}}{2} + \left( \cos \varphi I - i \sin \varphi \tilde{P} \right) \otimes \frac{I - \tilde{Q}}{2} \\
= e^{i \varphi \tilde{P}} \otimes \frac{I + \tilde{Q}}{2} + e^{-i \varphi \tilde{P}} \otimes \frac{I - \tilde{Q}}{2},
\label{eq:supp_core_induct}
\end{multline}
for every Pauli string $\tilde{P}$ and $\varphi \in \mathbb{R}$.
This proves the identity for $L=1$.
The inductive step follows immediately from multiplying the right-hand sides.

If $|\psi ^{\prime}\rangle = (I_{n-m} \otimes \tilde{Q}) |\psi\rangle$ is a state obtained by a one-time pairwise buffer exchange associated with $\tilde{Q}$,
and stored in the auxiliary buffers $B_k$,
the final state can be obtained by a linear combination of states:
\begin{equation}
\label{eq:supp_core_state}
\prod_{l=1}^{L} e^{i \varphi_l \tilde{P}_l \otimes \tilde{Q}} \, |\psi\rangle =
\frac{1}{\sqrt{2}}
\left(   \prod_{l=1}^{L} e^{i \varphi_l \tilde{P}_l} \otimes I_{m} \frac{|\psi\rangle + |\psi^{\prime}\rangle}{\sqrt{2}} +
 \prod_{l=1}^{L} e^{-i \varphi_l \tilde{P}_l} \otimes I_{m}
\frac{|\psi\rangle - |\psi^{\prime}\rangle}{\sqrt{2}} \right).
\end{equation}
The result of Eq.~\eqref{eq:supp_core_state} can be obtained in three steps:
(i)~For $|\psi\rangle$ distributed among partitions $[A_k]$,
$k = 0,1,\ldots, M -1$,
load the state $|\psi^{\prime}\rangle$ into the auxiliary buffers $[B_k]$.
(ii)~By taking an in-place linear combination,
load the state $\frac{|\psi\rangle + |\psi^{\prime}\rangle}{\sqrt{2}}$ into $[A_k]$
and $\frac{|\psi\rangle - |\psi^{\prime}\rangle}{\sqrt{2}}$ into $[B_k]$.
(iii)~Perform the action of $ \prod_{l=1}^{L} e^{\pm i \varphi_l \tilde{P}_l} \otimes I_{m}$ by all MPI tasks
independently and in parallel on lower qubits,
and store the linear combination of $[A_{k}]$ and $[B_{k}]$,
weighted by $\frac{1}{\sqrt{2}}$,
in $[A_{k}]$.

\subsubsection*{Pauli-string bit representation}

By choosing an efficient representation of Pauli strings in computer memory,
we can perform the action of $\tilde{P} = P_{1} \otimes P_{2} \otimes \ldots \otimes \tilde{P}_{n-m}$ on lower qubits at a rate comparable to the memory access time for all elements of the local array $[A_{k}]$.
Because by assumption $n \leq 64$,
we represent an $n$-qubit Pauli string as two 64-bit integer numbers:
$\tilde{P} \approx (p_{1}, p_{2})$,
$0 \leq p_{1}, p_{2} < 2^{64}$,
such that the $n$-th bit of $p_{1}$ is set if and only if the Pauli operator $P_{n}$ is \emph{not} diagonal,
$P_{n} \in \{ \sigma_{x}, \sigma_{y} \}$,
whereas the $n$-th bit of $p_{2}$ is set whenever $P_{n} \in \{ \sigma_{y}, \sigma_{z} \}$.
For example,
for the Pauli string $\sigma_{x} \otimes I \otimes \sigma_{y}$,
$p_{1} = (101)_{2} = 5$,
and $p_{2} = (100)_{2} = 4$.
With this representation,
for a computational basis state $|i\rangle$,
$i = 0, 1, \ldots, 2^{n} - 1$,
we obtain immediately that $\tilde{P} |i\rangle = \omega |j\rangle$,
where $j = p_{1} \, \oplus \, i$,
and the symbol $\oplus$ stands for `exclusive OR',
and $\omega$ is a 4th root of unity,
$\omega \in \{1, i, -1, -i\}$.
The value of $\omega$ is computed from the bit parity of
$p_{1} \, \wedge \, p_{2}$ (the number of $i$'s,
i.e. the number of $\sigma_{y}$ operators in the string $\tilde{P}$),
and $p_{2} \, \wedge \, i$ (the number of ``minuses'': when either $\sigma_{y}$ or $\sigma_{z}$ act on the $n$-th qubit state: $|1\rangle$).

\end{document}